\documentclass[12pt,a4paper]{article}
\usepackage{graphicx}
\usepackage[margin=2cm]{geometry}
\usepackage{amsmath,amssymb}
\usepackage{enumerate}
\usepackage[shortlabels]{enumitem}
\usepackage{subcaption}
\usepackage{bm}

\DeclareMathOperator*{\argmin}{argmin}

\providecommand{\gpuarray}{{\texttt{gpuArray}}}
\providecommand{\celegans}{\textit{C. elegans}}
\renewcommand{\epsilon}{\varepsilon}

\begin{document}
\title{Passively parallel regularized stokeslets}
\author{Meurig T. Gallagher\(^{1,2}\)\thanks{m.t.gallagher@bham.ac.uk} and David J. Smith\(^{3,2}\)\\[1em] \(^1\)Centre for Systems Modelling and Quantitative Biomedicine, \\ \(^2\)Institute for Metabolism and Systems Research,\\ \(^3\)School of Mathematics, \\University of Birmingham, Birmingham B15 2TT, UK}
\date{}

\maketitle

\begin{abstract}
\sloppy
    Stokes flow, discussed by G.G.\ Stokes in 1851, describes many microscopic biological flow phenomena, including cilia-driven transport and flagellar motility; the need to quantify and understand these flows has motivated decades of mathematical and computational research. Regularized stokeslet methods, which have been used and refined over the past twenty years, offer significant advantages in simplicity of implementation, with a recent modification based on nearest-neighbour interpolation providing significant improvements in efficiency and accuracy. Moreover this method can be implemented with the majority of the computation taking place through built-in linear algebra, entailing that state-of-the-art hardware and software developments in the latter, in particular multicore and GPU computing, can be exploited through minimal modifications (`passive parallelism') to existing MATLAB computer code. Hence, and with widely-available GPU hardware, significant improvements in the efficiency of the regularized stokeslet method can be obtained. The approach is demonstrated through computational experiments on three model biological flows: undulatory propulsion of multiple \celegans, simulation of progression and transport by multiple sperm in a geometrically confined region, and left-right symmetry breaking particle transport in the ventral node of the mouse embryo. In general an order-of-magnitude improvement in efficiency is observed. This development further widens the complexity of biological flow systems that are accessible without the need for extensive code development or specialist facilities.
\end{abstract}

\section{Introduction}
Stokes flow describes the fluid mechanics of a vast range of microscopic life, for example the motility and generation of feeding currents by microorganisms, and organ cleansing, gamete/embryo transport and developmental patterning in higher organisms. Typically flow and locomotion involve the action of individual or multiple slender organelles termed flagella and cilia, with the effects of cell surfaces, surrounding cavities and sometimes free surfaces playing crucial roles through hydrodynamic interaction. 

This biological relevance has motivated decades of research into what we now refer to as the Stokes flow equations, originally studied (with the inclusion of an unsteady term) by G.G.\ Stokes \cite{stokes1851}, which describe Newtonian fluid dynamics in the inertialess regime associated with microscopic length scales. The field is too broad to do justice to here, so we mention a few highlights, including Gray \& Hancock's study \cite{gray1955} of the mechanism of sea urchin sperm propulsion and associated slender body theory (ref.\ \cite{hancock1953}, see also \cite{burgers1938}, \cite{lighthill1976} and \cite{johnson1980}), Chwang \& Wu's work on helical propulsion (ref.\ \cite{chwang1971}, see also \cite{ramia1993} and \cite{lauga2006}), Blake's squirmer model \cite{blake1971squirmer} and method of images \cite{blake1971images} (see also \cite{ishimoto2013,pedley2016,liron1976,blake1996,ainley2008,cortez2015}), Pedley, Kessler and colleagues' studies of algal suspension dynamics and gyrotaxis \cite{pedley1987,hill2005}, the discovery of Stokes flow as the first left-right asymmetric event in vertebrate embryo development \cite{cartwright2004,brokaw2005} (see also \cite{cartwright2020}), Cortez and colleagues' development of the method of regularized stokeslets \cite{cortez2001,cortez2005}, and Goldstein and colleagues' discoveries on algal cilia fluid mechanics, from single cell to colony scales \cite{goldstein2015}. For further review see for example references \cite{lauga2009,montenegro2012,elgeti2015,goldstein2016,smith2019}. 

 Numerical techniques, such as the method of regularized stokeslets, have become increasingly valuable in progressing our understanding of biological Stokes flow; in this manuscript we will briefly review this approach, focussing on numerical discretisation techniques that achieve a balance between ease-of-implementation and computational efficiency. We will discuss how this method can be extended to utilise GPU processing capabilities. Algorithms vary greatly in how they improve in performance on parallel processing architectures in general, and GPU hardware in particular. We therefore assess what affect this has on the computational cost of the method by benchmarking against previous work \cite{gallagher2018,gallagher2020}, as well as providing new simulations of multiple undulatory swimmers and investigating whether sperm-like swimmers can be used to drive particle transport through an enclosed channel.  

\section{Stokes flow and stokeslets}
The Stokes flow equations describing the inertialess dynamics of an incompressible Newtonian fluid are
\begin{align}
    -\bm{\nabla}p+\mu\nabla^2\bm{u} & = \bm{0}, \\
    \bm{\nabla}\cdot\bm{u} & = 0, \label{eq:stokes}
\end{align}
where \(p=p(\bm{x},t)\) is pressure, \(\bm{u}=\bm{u}(\bm{x},t)\) is velocity, \(\bm{x}\) is position, \(t\) is time and the constant \(\mu\) is dynamic viscosity. These equations are typically accompanied by the no-slip, no-penetration condition \(\bm{u}(\bm{X},t)=\partial_t \bm{X}\) on boundary points \(\bm{X}\), along with \(\bm{u}\rightarrow 0\) or \(\bm{u}\sim\bm{U}_{\mathrm{ambient}}\) as \(|\bm{x}|\rightarrow \infty\) in exterior flows.

\begin{figure}
    \centering
    \includegraphics[width=0.9\textwidth]{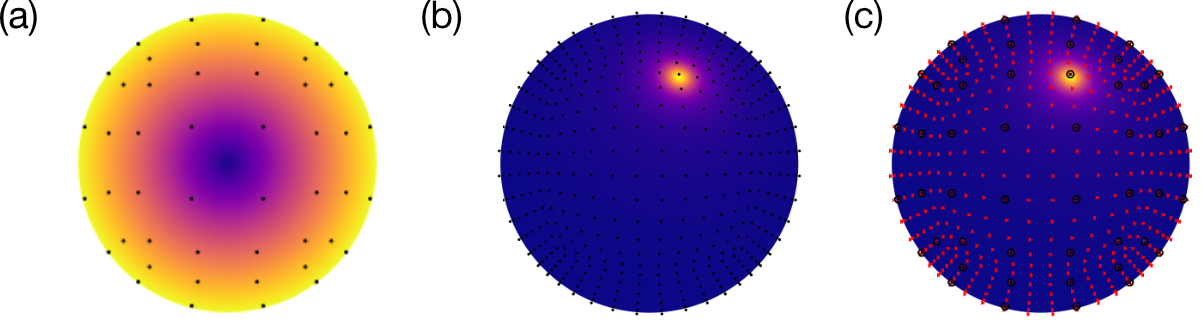}
    \caption{The problem of discretizing the regularized stokeslet boundary integral equations via quadrature rules, illustrated via the resistance problem for a rigid sphere, with associated coarse discretization illustrated via dots. (a) The traction field associated with pure rotation, with associated fine discretization illustrated via crosses. (b) The regularized stokeslet kernel with \(\epsilon=0.1\) for fixed stokeslet point \(\bm{y}\) and varying field point \(\bm{x}\). (c) The combination of coarse (black dot) and fine (red cross) discretizations as employed by the nearest neighbour regularized stokeslet method.}
    \label{fig:discretizations}
\end{figure}

The dynamic geometric complexity characterising many biological flows complicates both analytical and numerical solution approaches, however the linearity of equations~\eqref{eq:stokes} mean that solutions can be constructed by superposing `fundamental' solutions in the absence of a boundary driven by point forces, moments, sources/sinks and higher order derivatives. For example, Hancock's \cite{hancock1953} slender body theory was based on replacing the sperm flagellum with a line integral of point forces (which he termed \emph{stokeslets}) combined with appropriately-weighted source dipoles to account for the finite radius of the flagellum. Similarly, the flow due to a translating sphere in Stokes flow can be decomposed exactly as a point force and source dipole.

Another key feature of equations~\eqref{eq:stokes} is the absence of an explicit time derivative, which means that flow is instantaneously determined by the boundary conditions. This absence results in the famous scallop theorem \cite{purcell1977}, i.e.\ that a swimmer must undertake a time-irreversible motion in order to achieve a net translation (see also the earlier movie of G.I.\ Taylor \cite{taylor1985}); moreover a microscopic swimmer cannot `kick and glide', continuous motion requires continuous expenditure of energy. From a computational perspective, the absence of a time derivative entails that the flow problem itself does not require time-stepping, although the time-evolving transport of suspended particles or migration of swimmers of course does.

Focusing on fundamental solutions, the Stokes flow equations with a finite but spatially-concentrated force located at \(\bm{y}\) and pointing in the \(k\)-direction are
\begin{align}
    -\bm{\nabla}p+\mu\nabla^2\bm{u} + \delta(\bm{x}-\bm{y})\hat{\bm{e}}_k & = \bm{0}, \\
    \bm{\nabla}\cdot\bm{u} & = 0, \label{eq:stokesDelta}
\end{align}
where \(\delta(\bm{x})\) is the three-dimensional Dirac delta distribution and \(\hat{\bm{e}}_k\) is a unit basis vector pointing in the \(k\)-direction. Equations~\eqref{eq:stokesDelta} have solution \(\bm{u}=(8\pi\mu)^{-1}(S_{1k},S_{2k},S_{3k})\) and \(p=(4\pi)^{-1}P_k\), where
\begin{align}
    S_{jk}(\bm{x},\bm{y})& = \frac{\delta_{jk}}{|\bm{x}-\bm{y}|}+\frac{(x_j-y_j)(x_k-y_k)}{|\bm{x}-\bm{y}|^3},\\
    P_k(\bm{x},\bm{y})   & = \frac{x_k-y_k}{|\bm{x}-\bm{y}|^3}. \label{eq:stokeslet}
\end{align}
This \emph{Oseen tensor} \cite{oseen1927} or \emph{stokeslet} \cite{hancock1953} forms the basis for slender body theory \cite{burgers1938,hancock1953}, along with the boundary integral method \cite{youngren1975,phan1987,pozrikidis1992}. The boundary integral method enables the flow exterior to or bounded by a smooth surface \(B\) to be expressed as
\begin{equation}
    u_k(\bm{y})=-\frac{1}{8\pi\mu}\iint_B S_{jk}(\bm{x},\bm{y})f_j(\bm{x})dS_{\bm{x}} + \frac{1}{8\pi}\iint_{B} T_{jk\ell}(\bm{x},\bm{y})u_j(\bm{x})n_\ell(\bm{x})dS_{\bm{x}}, \label{eq:bie}
\end{equation}
where \(\bm{f}\) is the traction, i.e.\ the force per unit area exerted by the fluid on the surface, \(\bm{n}\) is the unit normal pointing into the fluid, and \(T_{jk\ell}\) is the stress tensor \(-P_k\delta_{j\ell}+\mu(\partial_\ell S_{jk}+\partial_j S_{\ell k}).\) The summation convection over repeated Cartesian indices is assumed above and throughout.

For problems satisfying \(\iint_B \bm{u}\cdot\bm{n} dS_{\bm{y}} = 0\), i.e.\ no change in volume, the second integral term (`double layer potential') in equation~\eqref{eq:bie} can be eliminated (see for example \cite{ishimoto2017}), motivating focus on the single layer boundary integral equation
\begin{equation}
    u_k(\bm{y})=-\frac{1}{8\pi\mu}\iint_B S_{jk}(\bm{x},\bm{y}) f_j(\bm{x}) dS_{\bm{x}}. \label{eq:slbie}
\end{equation}
Using the fact that \(S_{jk}(\bm{x},\bm{y})=S_{kj}(\bm{y},\bm{x})\) and relabelling, equation~\eqref{eq:slbie} may be rewritten \cite{pozrikidis1992}
\begin{equation}
    u_j(\bm{x}) = -\frac{1}{8\pi\mu}\iint_B S_{jk}(\bm{x},\bm{y}) f_k(\bm{y}) dS_{\bm{y}}. \label{eq:slbie2}
\end{equation}

The well-known advantages of the boundary integral are (1) reducing the computational domain from a 3D volume to a 2D surface (or possibly 1D line) \(B\), and (2) the natural treatment of open boundaries without the need to make artificial truncations, enabling excellent computational accuracy and efficiency. For example, Phan-Thien et al.\ were able to model propulsion of a physiologically-shaped bull sperm cell in the mid-1980s, making use of machines with the order of 1 MB memory \cite{phan1987}.

The standard numerical approach to the solution of equation~\eqref{eq:slbie2} is to discretize the traction as
\begin{equation}
    \bm{f}(\bm{y}) \approx \sum_{n=1}^N \bm{f}[n] \phi_n(\bm{y}), \label{eq:basis}
\end{equation}
where \(\bm{f}[1],\ldots,\bm{f}[N]\) are vector constants and \(\phi_1(\bm{y}),\ldots,\phi_N(\bm{y})\) are basis functions (see Figure~\ref{fig:discretizations} for a sketch of the discretisation over a rigid sphere). In the simplest case, the latter may be piecewise constant functions defined on a partition \(B=B_1\cup \ldots \cup B_N\) via \(\phi_n(\bm{y})=1\) for \(\bm{y}\in B_n\) and \(\phi_n(\bm{y})=0\) otherwise. The semi-discrete numerical problem then reads
\begin{equation}
    u_j(\bm{x}) = - \frac{1}{8\pi\mu} \sum_{n=1}^N \left(\iint_{B_n} S_{jk}(\bm{x},\bm{y}) dS_{\bm{y}}\right) f_k[n]. \label{eq:dbi}
\end{equation}
The resistance problem (prescribed velocity \(\bm{u}\) on the surface, unknown traction \(\bm{f}\)) can be solved as follows: equation~\eqref{eq:dbi} can be converted to a \(3N\times 3N\) linear system by applying point collocation, i.e.\ \(\bm{x}=\bm{x}[m]\) for \(m=1,\ldots, N\) where \(\bm{x}[m]\) is the centroid of element \(B_m\). As discussed below, the swimming problem is a minor modification of the resistance problem involving the introduction of an unknown rigid body motion and additional force and moment balance equations.

However, scientists who are not computational specialists may encounter two challenges in implementing the above approach:
\begin{enumerate}
    \item The generation of a \emph{mesh}, i.e.\ the connected, non-overlapping partition \(B_1,\ldots, B_N\) of what may be a complex geometry. Typically this mesh includes a set of vertices, a connectivity table associating each element \(B_n\) to 3 or 4 vertices, a mapping between local coordinates \((\xi,\eta)\) in each element and global coordinates \(\bm{y}\) and surface metric \(dS_{\bm{y}}\).
    \item The evaluation of the (weakly) singular integrals occurring in equation~\eqref{eq:dbi} when \(\bm{x}=\bm{x}[m]\) and \(n=m\). These integrals involve terms which behave as \(r^{-1}\) as \(r\rightarrow 0\). A related difficulty occurs `downstream' of the problem for the traction if subsequent computation of the velocity \(\bm{u}(\bm{x})\) at a fluid point \(\bm{x}\) close to \(B\) is required.
\end{enumerate}

\section{Regularized stokeslets}
Issues (i) and (ii) are certainly not insurmountable, and codes such as the Fortran library BEMLIB \cite{pozrikidis2002} provide considerable assistance. However, given the scientific breadth of (often experimentally-focused or multidisciplinary) researchers working in biological fluid dynamics, there is great value in methods which avoid or alleviate them. One appealing strategy is to \emph{regularize} the singularity occurring so that \(\tilde{S}_{jk}(\bm{x},\bm{x})<\infty\), which immediately solves issue (ii). Additionally issue (i) can then be side-stepped by simply covering \(B\) with a set of points \(\{\bm{x}[1],\ldots,\bm{x}[N]\}\) (without the need for a connectivity table or local coordinates), and discretizing the integral directly with a quadrature rule,
\begin{equation}
    \iint_B \tilde{S}_{jk}(\bm{x},\bm{y}) f_k(\bm{y}) dS_{\bm{y}} \approx \sum_{n=1}^N \tilde{S}_{jk}(\bm{x},\bm{x}[n])F_k[n], 
\end{equation}
where \(F_k[n]:=f_k(\bm{x}[n])dS(\bm{x}[n])\), i.e. absorbing the quadrature weight and metric into the unknown force.  The linear system for the resistance problem then becomes,
\begin{equation}
    u_j(\bm{x}[m]) = -\frac{1}{8\pi\mu}\sum_{n=1}^N \tilde{S}_{jk}(\bm{x}[m],\bm{x}[n]) F_k[n].\label{eq:nystrom}
\end{equation}
Equation~\eqref{eq:nystrom} is referred to as a Nystr\"{o}m discretization of the integral equation \cite{nystrom1930}.

\begin{figure}
    \centering
    \includegraphics[width=0.9\textwidth]{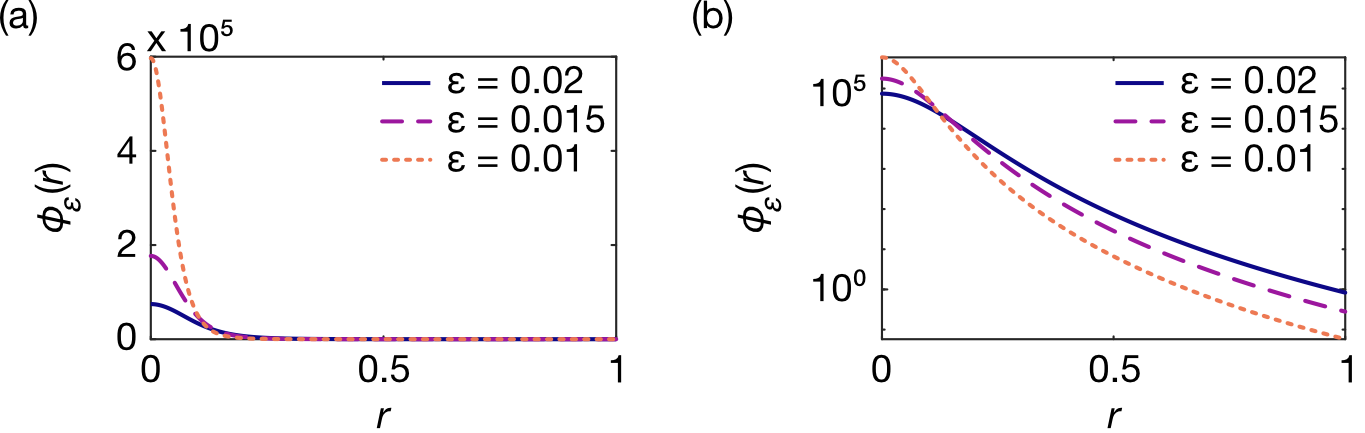}
    \caption{The spatially-smoothed approximation \(\phi_\epsilon\) to the 3D Dirac delta function of Cortez et al. \cite{cortez2005}, plotted in spherical polar coordinates, for values \(\epsilon=0.02, 0.015, 0.01\) on (a) linear and (b) log-linear axes.}
    \label{fig:blobs}
\end{figure}

There are clearly many ways to regularize the kernel, for example replacing \(|\bm{x}-\bm{y}|^{-1}\) with \((|\bm{x}-\bm{y}|^2+\epsilon^2)^{-1/2}\) for some small parameter \(\epsilon > 0\). However, this ad hoc approach has the unwanted side-effect of producing potentially unphysical ``solutions'' that no longer satisfy the original equations, for example by violating the incompressibility condition \(\bm{\nabla}\cdot \bm{u}=0\). Cortez and colleagues \cite{cortez2001, cortez2005} proposed instead to start by regularizing the point force, i.e.\ to consider the exact solution to the Stokes flow equations with spatially-smoothed point forces,
\begin{align}
    -\bm{\nabla}p+\mu\nabla^2\bm{u} + \phi_\epsilon(\bm{x}-\bm{y})\hat{\bm{e}}_k & = \bm{0}, \\
    \bm{\nabla}\cdot\bm{u} & = 0. \label{eq:stokesReg}
\end{align}
where \(\phi_\epsilon(\bm{x})\) is a family of ``blob'' functions which approximates \(\delta(\bm{x})\) as \(\epsilon\rightarrow 0\). The particular choice
\begin{equation}
    \phi_\epsilon(\bm{x})=\frac{15\epsilon^4}{8\pi(|\bm{x}|^2+\epsilon^2)^{7/2}}, \label{eq:blob}
\end{equation}
is plotted in Figure~\ref{fig:blobs}. This choice leads to the regularized stokeslet solutions
\begin{align}
    P_k^\epsilon(\bm{x},\bm{y}) & = \frac{x_k}{(|\bm{x}|^2+\epsilon^2)^{5/2}}(2|\bm{x}|^2+5\epsilon^2),\\
    S_{jk}^\epsilon(\bm{x},\bm{y}) & = \frac{\delta_{jk}(|\bm{x}|^2+2\epsilon^2)+x_j x_k}{(|\bm{x}|^2+\epsilon^2)^{3/2}}.
\end{align}

It can be seen that \(P_k^\epsilon(\bm{x})\sim P_k(\bm{x})\) and \(S_{jk}^\epsilon(\bm{x},\bm{y})\) as \(\epsilon\rightarrow 0\); moreover the corresponding single layer boundary integral equation is
\begin{equation}
    u_j(\bm{x}) = -\frac{1}{8\pi\mu}\iint_B S_{jk}^\epsilon(\bm{x},\bm{y}) f_k(\bm{y}) dS_{\bm{y}} + O(\epsilon^p), \label{eq:regBie}
\end{equation}
where \(p=1\) for \(\bm{x}\) on or near \(B\) and \(p=2\) otherwise \cite{cortez2005}. Alternative forms for the blob function have been derived \cite{cortez2015} to improve the order of the error, however the form~\eqref{eq:blob} is by far the most commonly-used. The \(O(\epsilon)\) regularization error associated with boundary collocation is inherited by the resistance problem for finding the traction associated with a given rigid body motion, and the swimming problem for finding the traction and translation/rotation resulting from a given boundary deformation and force/moment balance. It is distinct from the errors associated with discretization of the traction and numerical quadrature which we will consider shortly. 

Regularization enables a particularly convenient discretization of the single layer boundary integral equation; this simplicity however comes at a cost. The Nystr\"{o}m method \eqref{eq:nystrom} corresponds to using identical discretizations for the traction \(f_k(\bm{y})\) and \(S_{jk}(\bm{x},\bm{y})\) when considered as functions of position \(\bm{y}\in B\) (the latter for fixed \(\bm{x}\)), despite the fact that the stokeslet kernel varies much more rapidly than the traction, as shown in figure~\ref{fig:discretizations}, the variability becoming more rapid as \(\epsilon\) is reduced (Figure~\ref{fig:blobs}). The number of degrees of freedom of the system \(3N\) and hence the \(3N\times 3N\) matrix size is tied to the discretization for the traction, and so reducing the regularization error via reducing \(\epsilon\) entails rapid growth in memory and computational requirements associated with system assembly and solution. Cortez et al.\ \cite{cortez2005} initially suggested a quadrature error \(O(h^2/\epsilon^3)\) where \(h\) is the discretization length; more recently \cite{gallagher2019} this estimate has been improved to \(O(h^2/\epsilon)+O(P\epsilon^{-1/P} h^{1-P})\) for any integer \(P>3\). 

As described in equations~\eqref{eq:basis},~\eqref{eq:dbi}, boundary element methods separate out the traction and quadrature discretization by expanding the traction in terms of basis functions, leaving the stokeslet integrals to be evaluated by the most appropriate means (e.g.\ adaptive quadrature) without affecting the number of degrees of freedom of the system. Alongside being the standard method for the classical boundary integral equation, this approach has been employed in the context of \emph{regularized} stokeslets to model cilia-driven flow \cite{smith2012,sampaio2014}, autophoretic swimmers \cite{montenegro2015} and to explore the evolution of bacterial morphology \cite{schuech2019}. These studies would have been challenging or practically impossible via the Nystr\"{o}m method, which computational experiments suggest would have required above 3 orders of magnitude greater memory and processing time \cite{smith2009}.
A related idea of pre-integrating line distributions of regularized stokeslets to model cilia and flagella, which is closely related to slender body theory \cite{smith2009} has recently been extended to higher order basis functions \cite{cortez2018,walker2019}, and to model flagellar elastohydrodynamics \cite{hall2019}. However, problems involving surface integrals require true mesh generation and as such despite being suggested over 10 years ago \cite{smith2009}, the `element' approach has not been very widely adopted in comparison with the Nystr\"{o}m method for regularized stokeslets.

With the aim of preserving the meshlessness of the Nystr\"{o}m discretization but decoupling the traction discretization from the numerical quadrature, we recently \cite{smith2018} suggested an alternative approach based on an idea from meshless interpolation. The concept is to use two point cloud discretizations, one `coarse force' set \(\{\bm{x}[1],\ldots,\bm{x}[N]\}\) which is sufficient to capture the variation of the traction, and which dictates the size of the linear system, and another finer quadrature discretization set \(\{\bm{X}[1],\ldots,\bm{X}[Q]\}\) which has sufficient resolution to capture the rapidly-varying kernel \(S_{jk}^\epsilon(\bm{x},\bm{y})\) for \(\bm{y}\) in the vicinity of \(\bm{x}\), as in Figure~\ref{fig:discretizations}c. The force at a quadrature point \(\bm{f}(\bm{X}[q])dS(\bm{X}[q])\) is then approximated by its value at the nearest point on the coarse force set, \(\bm{f}(\bm{x}[\hat{n}])dS(\bm{x}[\hat{n}])\), where the index \(\hat{n}\) is given by
\begin{equation}
    \hat{n}=\argmin_{n=1,\ldots,N}{|\bm{x}(n)-\bm{X}(q)|}.
\end{equation}
Denoting \(\nu[q,\hat{n}]=1\) and \(\nu[q,n]=0\) for \(n\not = \hat{n}\), we may write
\begin{align}
    u_j(\bm{x}[m]) & \approx \frac{1}{8\pi\mu}\sum_{q=1}^Q S_{jk}^\epsilon(\bm{x}[m],\bm{X}[q])f_k(\bm{X}[q])dS(\bm{X}[q]) \\
                   & \approx \frac{1}{8\pi\mu}\sum_{q=1}^Q S_{jk}^\epsilon(\bm{x}[m],\bm{X}[q])\sum_{n=1}^N \nu[q,n] f_k(\bm{x}[n])dS(\bm{x}[n]) \\
                   & = \frac{1}{8\pi\mu}\sum_{n=1}\left(\sum_{q=1}^Q S_{jk}^\epsilon(\bm{x}[m],\bm{X}[q]) \nu[q,n]\right)F_k[n], \label{eq:nnrsbie}
\end{align}
where \(F_k[n]:=f_k(\bm{x}[n])dS(\bm{x}[n])\).

Defining \(h_f\) to be the length scale associated with the traction points and \(h_q\) to be the length scale associated with the quadrature points, equation~\eqref{eq:nnrsbie} has a regularization error \(O(\epsilon)\) and traction discretization error \(O(h_f)\). The quadrature error depends on whether the traction points are `contained' (i.e.\ within distance \(O(\epsilon)\) or closer) in the quadrature set. Associated to each traction point which is contained in the quadrature set is an error \(O(\epsilon^{-1}h_q^2)+O(P\epsilon^{-1/P}h_q^{1-1/P})\) for all integer \(P>3\), the former term being dominant \cite{gallagher2019}. Associated to each traction point which is no closer than distance \(\delta\gg \epsilon >0\) to the quadrature points is an error \(O(h_q^3\delta^{-2})+O(Ph_q^{1-2/P})\) \cite{gallagher2019}. In the latter case there is no asymptotic dependence on \(\epsilon\) in the limit \(\epsilon\rightarrow 0\).

The ``NEAREST'' method \eqref{eq:nnrsbie} is not intended to compete directly with boundary element methods, still less techniques based on treecodes and fast multipole methods \cite{rostami2016,rostami2019,wang2018}, rather it is aimed at providing a simple and accessible meshfree method for non-specialists that is capable of dealing with problems of moderate complexity, with relatively standard workstation hardware. As an example, the resistance matrix (forces and moments due to the translation and rotation of a sphere), with regularization parameter \(\epsilon=0.01\) can be calculated to within 2\% error with \(N=864\) and \(Q=3456\), taking 6 seconds on a notebook computer. The same computation with the Nystr\"{o}m method (\(N=Q=3456\)) yielded a 5\% error and required 350 seconds. Other early applications of the nearest neighbour discretization by our group have included the diffusion tensor of a macromolecular structure \cite{smith2018}, cell motility \cite{gallagher2018} and embryonic nodal flow \cite{gallagher2020}.

A further feature of this method is that equation~\eqref{eq:nnrsbie} can be viewed as the product of a (dense) \(3N\times 3Q\) stokeslet matrix with a \(3Q\times 3N\) (sparse) nearest neighbour matrix and a \(3N\times 1\) column vector of tractions, which naturally lends itself to implementation in numerical linear algebra software such as MATLAB or GNU Octave in terms of matrix-matrix and matrix-vector operations. Limited use of an iterated loop to handle the `Q' dimension in block prevents memory overflow in in this respect. The use of `under the hood' numerical linear algebra enables ongoing progress in hardware and software optimizations -- particularly those involving multicore and graphical processing unit (GPU) developments -- to be exploited `passively', i.e.\ with minimal changes to the code. Careful investigation of the effect of parallelisation is warranted; algorithms may differ by orders of magnitude in their performance on parallel architectures, and on GPU hardware in particular. It is important to note that inherently serial algorithms will yield at best no improvements on parallel hardware, and at worst can exhibit worse performance. We conclude this review by presenting some experiments along these lines; it will be found that the use of a modest compute-GPU in constructing and solving swimming-type problems can lead to a reduction in the required computational time that can be in excess of an order of magnitude when using the nearest neighbour regularized stokeslet method.

\section{Parallelising NEAREST}

In a recent article \cite{gallagher2018}, we showed how the trajectories of multiple flagellated swimmers with specified beat patterns can be calculated through formulating and solving an initial-value problem (IVP) of the form \(\dot{Y} = F(Y,t)\), where
\begin{equation*}
    Y(t) = \begin{bmatrix}
            \mathbf{X}_0(t)\\
            \mathbf{B}(t)
        \end{bmatrix},
\end{equation*}
with \(\mathbf{X}_0(t)\) being the position of the swimmers at time \(t\), and \(\mathbf{B}(t)\) the matrix of basis vectors describing the body-frame of the swimmers. The rates \(\dot{\mathbf{X}}_0(t)\), and \(\dot{\mathbf{B}}(t)\) are determined from the linear system for the swimmers' translational velocity \(\mathbf{U}\) (3 scalar unknowns per swimmer), angular velocity \(\mathbf{\Omega}\) (3 scalar unknowns per swimmer), and tractions \(\mathbf{f}[\cdot]\) (\(3N\) scalar unknowns per swimmer). This formulation allows for the use of built-in IVP solvers such as MATLAB's {\texttt{ode45}}, or GNU Octave's {\texttt{lsode}}. Approximately \(90\% - 99\%\) of the computational time needed to solve such a problem using NEAREST is due to a combination, at each time step, of:
\begin{enumerate}
    \item Construction of the stokeslet matrix multiplied by the nearest neighbour matrix (\(A_s\)), such as in~\eqref{eq:nnrsbie};
    \item Construction of the matrix (\(A\)) from pre-calculated matrix blocks;
    \item Solution of the linear system.
\end{enumerate}

% Not sure if this needs a little tweak
The construction of the matrix \(A_s\) can itself be decomposed into two steps: the calculation of the regularized stokeslet kernel at each force node with respect to each quadrature node (\(3N M\times 3Q M\) calculations for \(M\) swimmers), and, then, the multiplying of the stokeslet matrix with the nearest neighbour matrix \(\nu[q,n]\). This is exactly the type of problem that is suited to GPU optimisation; a large number of small calculations can take advantage of the large number of processing cores available on modern GPUs, with relatively modest hardware enabling thousands of calculations to be performed simultaneously. Similarly, it is well known that there are significant gains to be made when using a GPU for large matrix construction \cite{dziekonski2012finite}, and with the NEAREST method we are able to exploit commercially optimised algorithms for solving linear systems on the GPU, such as the MATLAB ``\(\backslash\)'' command.

With the aim of maintaining the simplicity of the original method we take a \textit{passively parallel} approach whereby we do not rewrite the entire method to fully optimise GPU use, but make minimal changes that nonetheless bring an order of magnitude performance improvement. For detailed description of the previously-published code and algorithms, see refs.\ \cite{smith2018,gallagher2018}; the entirety of the code changes to make use of an available compute GPU are as follows:
\begin{enumerate}
    \item To construct the matrix \(A_s\), the vectors containing force and quadrature points (\(x\)~and~\(X\) respectively) are `cast' to the GPU via the commands
        \begin{equation*}
            x = \gpuarray\left(x\right),\quad X = \gpuarray\left(X\right).
        \end{equation*}
        There is a small amount of computational overhead in transferring data between the CPU and GPU, but many MATLAB operators are `overloaded' to work on both standard and GPU arrays without additional intervention by the user. Due to the use of \(x\) and \(X\) as \gpuarray s, the constructed matrix \(A_s\) is itself a \gpuarray, and therefore the large matrix \(A\) is constructed on the GPU, again with no additional user intervention.
        
    \item The MATLAB ``\(\backslash\)'' command for solving the linear system works without modification on a \gpuarray, outputting a \gpuarray\ solution vector \(\dot{F}\). Currently the MATLAB built-in ODE solvers (such as {\texttt{ode45}}) are not currently GPU optimised, and so the solution must be `gathered' from the GPU back to the CPU via the command
        \begin{equation*}
            \dot{F} = {\texttt{gather}}(\dot{F}).
        \end{equation*}
\end{enumerate}

In what follows we will investigate the effect that these minimal user changes can have on the run time of simulations, including an assessment of the savings for each of the three major costs (assembling \(A_s\), assembling \(A\), linear solver), and how these changes impact the time required to solve `real-world' problems
based on previous work of tracking \(9\) sperm swimming between two solid plate boundaries (an extension of \cite{gallagher2018}), and of tracking particle deposition in the \textit{euciliated} embryonic node of the mouse. Once we have established how the use of GPU processing can improve existing computations, we will apply these modifications to two new sets of simulations involving 1. an array of \(25\) individually modelled swimming \textit{Caenorhabditis (C.) elegans}, and the problem of particle transport by an array of sperm swimming between two solid plate boundaries.

\subsection{Benchmarking}
\label{sec:benchmarks}

To benchmark the improvements when using a GPU to accelerate NEAREST, we make use of three machines:
\begin{enumerate}[start=1,label={{\bfseries M\arabic*}:},leftmargin=*]
    \item Lenovo Thinkstation, with 2x Intel Xeon 4116 CPUs (each with 12 cores), 128GB DDR4 RAM, and an NVIDIA Quadro RTX 5000 GPU (3072 parallel-processing cores, 16 GB GDDR6 RAM);
    \item Lenovo NeXtScale server \textit{BEAR HPC} (Birmingham Environment for Academic Research High Performance Computing) processing unit (CPU), using 1x Intel Xeon E5-2640 (10 cores) and 64GB DDR4 RAM;
    \item Lenovo NeXtScale server \textit{BEAR HPC (GPU)}, using 1x Intel Xeon E5-2640 CPU (10 cores), 64GB DDR4 RAM, and an NVIDIA Tesla P100 GPU (3584 parallel-processing cores, 12 GB HBM2 RAM).
\end{enumerate}
Comparisons will be made between calculations on M1 (with and without GPU acceleration), and between M2 (CPU) and M3 (GPU). Although M2 and M3 form part of an HPC cluster, each calculation will be limited to a single computational node per simulation. 

For benchmarking we take the swimming problem of a single model sperm from~\cite{gallagher2018}. The model sperm comprises a tail of length \(L = 50~\mu\)m, discretised as a line of stokeslets, and an ellipsoidal head with semi-axes of length \(2~\mu\)m, \(1.6~\mu\)m, and \(1~\mu\)m, and is nondimensionalised with respect to the length-scale \(L\). The beat pattern follows the planar activated beat of Dresdner and Katz~\cite{dresdner1981relationships}. We discretize the model sperm tail using \(N_t = 100\) and \(Q_t = 400\), and an increasing number of traction points on the head \(N_h\) with \(Q_h = 4N_h\), to obtain a problem in terms of \(3\left(N_h + 100\right)\) scalar degrees of freedom (sDOF), and corresponding \(3\left(4N_h + 400\right)\) quadrature points.

\begin{figure}[tp]
    \centering
    \includegraphics[width=\textwidth]{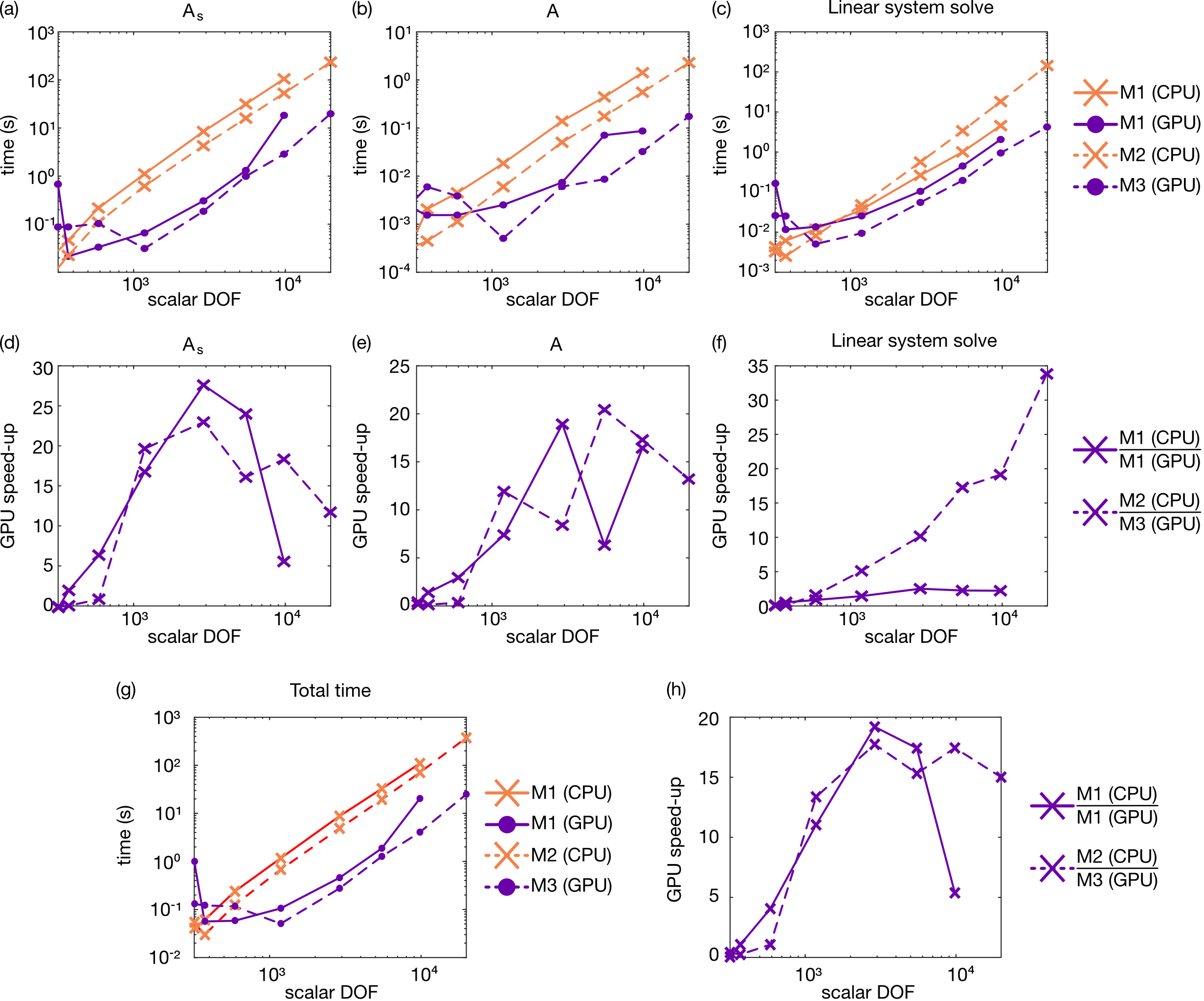}
    \caption{Comparison of the computational time for constructing the matrices \(A_s\) and \(A\), and solving the linear system for the problem of a single sperm swimming. (a) -- (c) The time taken for M1 (CPU), M1 (GPU), M2 and M3 (lower is better). (d) -- (f) The relative speed increase when incorporating the GPU for M1, and when using M3 instead of M2 (higher is better). (g) The total time to solve a single iteration for the single sperm on M1 (CPU), M1 (GPU) (lower is better), M2 and M3. (h) The relative speed increase when incorporating the GPU for M1 and M3 instead of M2 (higher is better).}
    \label{fig:mat_comparison}
\end{figure}

In Figure~\ref{fig:mat_comparison} we show the time required for calculation of the model swimming sperm at the first time point \(t = 0\). In these figures, the times shown are the mean time calculated over \(100\) runs. Figure~\ref{fig:mat_comparison}a--c shows the total time taken for constructing \(A_s\), \(A\) and for solving the linear system, with the relative increases (M1 (CPU)) / (M1 (GPU)) and M2 / M3 shown in Figure~\ref{fig:mat_comparison}d--f. In each case we see that for small numbers of sDOF (\(<500\)) the CPU calculations show a modest performance increase over those with GPU acceleration. When increasing the number of sDOF, we see greater than an order of magnitude speed up for \(A_s\) (maximum gains are a factor of 29 (M1) and 23 (M2/M3)), \(A\) (maximum gains are a factor of 19 (M1) and 20 (M2/M3)), and for solving the linear system on M2/M3 (34 times faster). When solving the linear system on M1 for \(>500\) sDOF, we see a factor of 2 increase when using the GPU as opposed to the CPU. This reduced increase is due to the availability of 24 CPU cores on M1, allowing for significant performance increases over M2. For very large numbers of sDOF (\(> 10^4\)) there is a slight reduction in computational gains when using the GPU machines. This is due to the limited amount of GPU memory available, meaning additional overhead in transferring data between the CPU and GPU. Figure~\ref{fig:mat_comparison}g,h shows the total time taken to construct and solve the swimming problem for the first time step \(t = 0\). Here same pattern occurs, with an order of magnitude speed up when using the GPU for problems with \(>10^3\) sDOF, with a maximum 21 times increase for M1, and 18 times increase for M2/M3.

\begin{figure}[t]
    \centering 
    \begin{subfigure}{\textwidth}
        \includegraphics[width=0.8\textwidth]{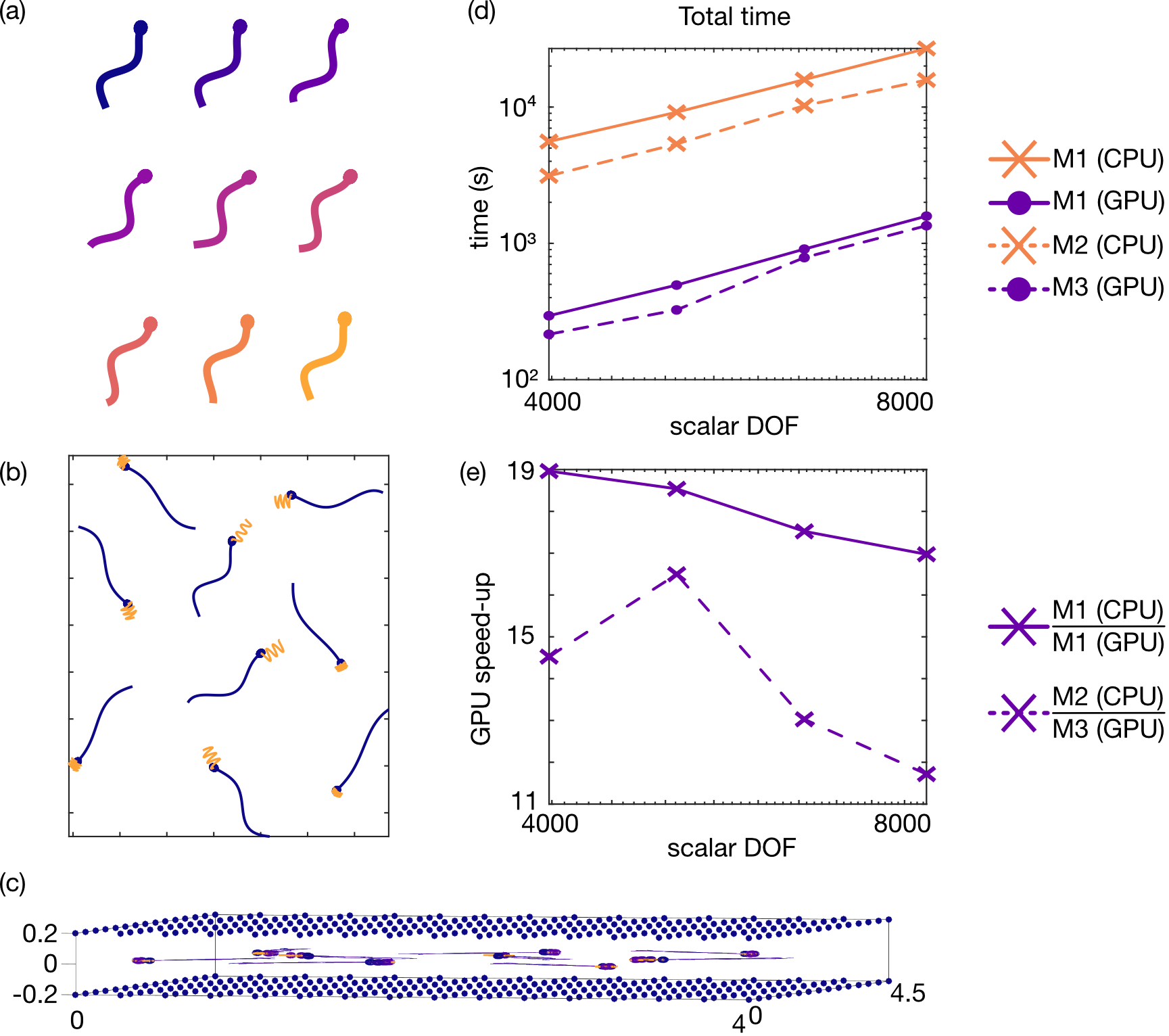}
        \phantomsubcaption{}
        \label{fig:manysperm_beat}
    \end{subfigure}
    \begin{subfigure}{0\textwidth}
        \phantomsubcaption{}
        \label{fig:manysperm_tracks}
    \end{subfigure}
    \begin{subfigure}{0\textwidth}
        \phantomsubcaption{}
        \label{fig:manysperm_side}
    \end{subfigure}
    \begin{subfigure}{0\textwidth}
        \phantomsubcaption{}
        \label{fig:manysperm_time}
    \end{subfigure}
    \begin{subfigure}{0\textwidth}
        \phantomsubcaption{}
        \label{fig:manysperm_speedup}
    \end{subfigure}
    \caption{Simulation of an array of 9 model sperm for three beat cycles. (a) The model beat pattern in time. (b) The position of each swimmer at \(t=0\) is shown, with the paths traced out by the leading point as they swim overlain. (c) Side-on view of swimmers showing location of plate boundaries. (d) The total simulation time required on each of M1 (CPU), M2 (GPU), M3 (CPU) and M4 (GPU) (lower is better). (e) The relative speed when using a GPU compared to CPU on M1, and M3 compared to M2 (higher is better).}
    \label{fig:manysperm}
\end{figure}

\subsection{Multiple swimming sperm between parallel plates}
\label{sec:multisperm}

In the first of our `real-world' tests, we assess the performance of the GPU parallelised method for a set of \(9\) sperm swimming between two parallel plates, using the knowledge that regularized stokeslet methods enable solid boundaries to be taken into account via the inclusion of additional surface distributions. In the present case we consider the situation of two parallel plane walls approximating a microscope slide and cover slip. Between these, we simulate an array of 9 model sperm, again using the model of Dresdner and Katz~\cite{dresdner1981relationships}, initialised with varying wave number \(k\in\left[2\pi,4\pi\right]\) and phase \(\phi\in\left[0,2\pi\right]\), and solved for three beat cycles. Boundaries are included as a pair of parallel rectangular plates \(4\times 4.5\) flagellar lengths (equivalently \(180\times 202.5~\mu\)m), separated by a distance of \(0.4\) flagellar lengths (equivalently \(18~\mu\)m), with the sperm swimming in the plane a distance of \(0.2\) flagellar lengths from each plate. The characteristic beat pattern of one of these swimmers is shown in Figure~\ref{fig:manysperm_beat}, with the position of each swimmer at time \(t = 0\), and the tracks traced out over three beat cycles shown in Figure~\ref{fig:manysperm_tracks}.  The boundaries (as shown in Figure~\ref{fig:manysperm_side}) are discretised with a total of \(1440\) sDOF (and corresponding \(7680\) vector quadrature points), with the sperm tails discretised using \(120\) sDOF (and corresponding \(160\) vector quadrature points), and the heads are discretised with an increasing number of sDOF (\(162\)--\(648\) per swimmer). For the range of sDOF shown in Figures~\ref{fig:manysperm_time}, \ref{fig:manysperm_speedup}, we see in excess of an order of magnitude speed increase when using the GPU accelerated machines, with a maximal \(19\ \times\) speed-up when using the GPU on M1 with \(3978\) sDOF.

\begin{figure}[t]
    \centering
    \begin{subfigure}{\textwidth}
        \includegraphics[width=0.8\textwidth]{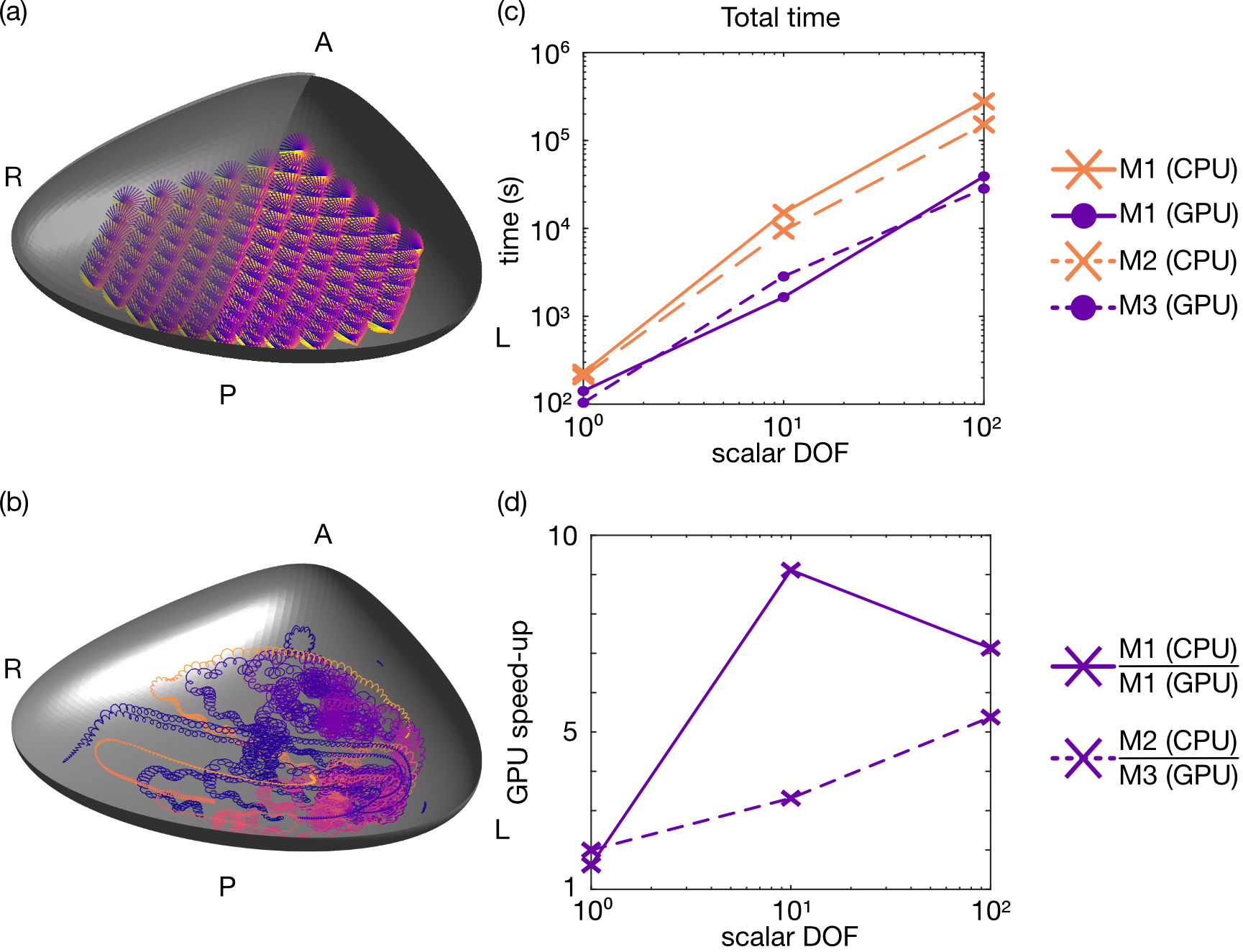}
        \phantomsubcaption{}
        \label{fig:node_sketch}
    \end{subfigure}
    \begin{subfigure}{0\textwidth}
        \phantomsubcaption{}
        \label{fig:node_tracks}
    \end{subfigure}
    \begin{subfigure}{0\textwidth}
        \phantomsubcaption{}
        \label{fig:node_time}
    \end{subfigure}
    \begin{subfigure}{0\textwidth}
        \phantomsubcaption{}
        \label{fig:node_speedup}
    \end{subfigure}
    \caption{Particle tracking in the \textit{euciliated} mouse node. (a) A sketch of the mouse node geometry with overlaying Reichert's membrane and 112 beating cilia. (b) Paths traced out by 10 particles after 1000 beats. (c) The total simulation time required on each of M1 (CPU), M2 (GPU), M3 (CPU) and M4 (GPU) (lower is better). (d) The relative speed when using a GPU compared to CPU on M1, and M3 compared to M2 (higher is better).}
    \label{fig:node}
\end{figure}

\subsection{Particle tracking in the \textit{euciliated} mouse node}
\label{sec:node}

The flow driven by beating cilia in a fluid-filled structure called the embryonic node is the initiator of symmetry breaking in early development \cite{smith2012,sampaio2014,smith2019}, although the mechanism by which this flow is converted into anatomical asymmetry is still to be understood \cite{cartwright2020}. The role of morphogen-bearing vesicles is believed to play a central role. 

To investigate this problem, we recently modelled the enclosed embryonic node of the mouse using NEAREST \cite{gallagher2020}, with a biologically realistic \(112\) beating cilia, with particular focus on how the calculated flow can transport particles potentially needed for chemical signalling. The large number of beating cilia, and non-planar boundaries in this problem, means that long-timescale computational simulation of particle transport is particularly challenging for most existing methods. 

Taking the forces calculated for the \(39,979\) sDOF from the previous work \cite{gallagher2020}, we calculate the time required to track a set of \(1\), \(10\), and \(100\) particles for \(1000\) beat cycles of the \(112\) cilia. A sketch of the \textit{euciliated} mouse node is shown in Figure~\ref{fig:node_sketch}, with the paths of \(10\) particles shown in Figure~\ref{fig:node_tracks}, time requirements in Figure~\ref{fig:node_time} and relative speed increase in Figure~\ref{fig:node_speedup}. We see in the tracks the characteristic leftward-moving, ``loopy drift'' of particles as they follow the flow generated by the array of beating cilia. The ability to model and perturb a physiologically-accurate node, with particle transport over a long time, alongside the mechanical stresses produced by cilia movement, may enable us to probe more deeply the flow conversion mechanism that has proved to be so elusive, and moreover to assess the diverse morphologies observed in different species. 

For the GPU accelerated calculations, the size of the system requires significant data transfer between the GPU and the CPU, and so the relative speed increases are not as significant -- up to almost an order of magnitude on M1, and half an order of magnitude when using M3 rather than M2. However the real-world implication of this speed increase is significant; when solving for a system of \(100\) particles, the time required is a little over \(77\)~hours when using M1 (CPU), but less than \(11\)~hours on M1 (GPU).

\begin{figure}[t]
    \centering
    \begin{subfigure}{\textwidth}
        \includegraphics[width=0.9\textwidth]{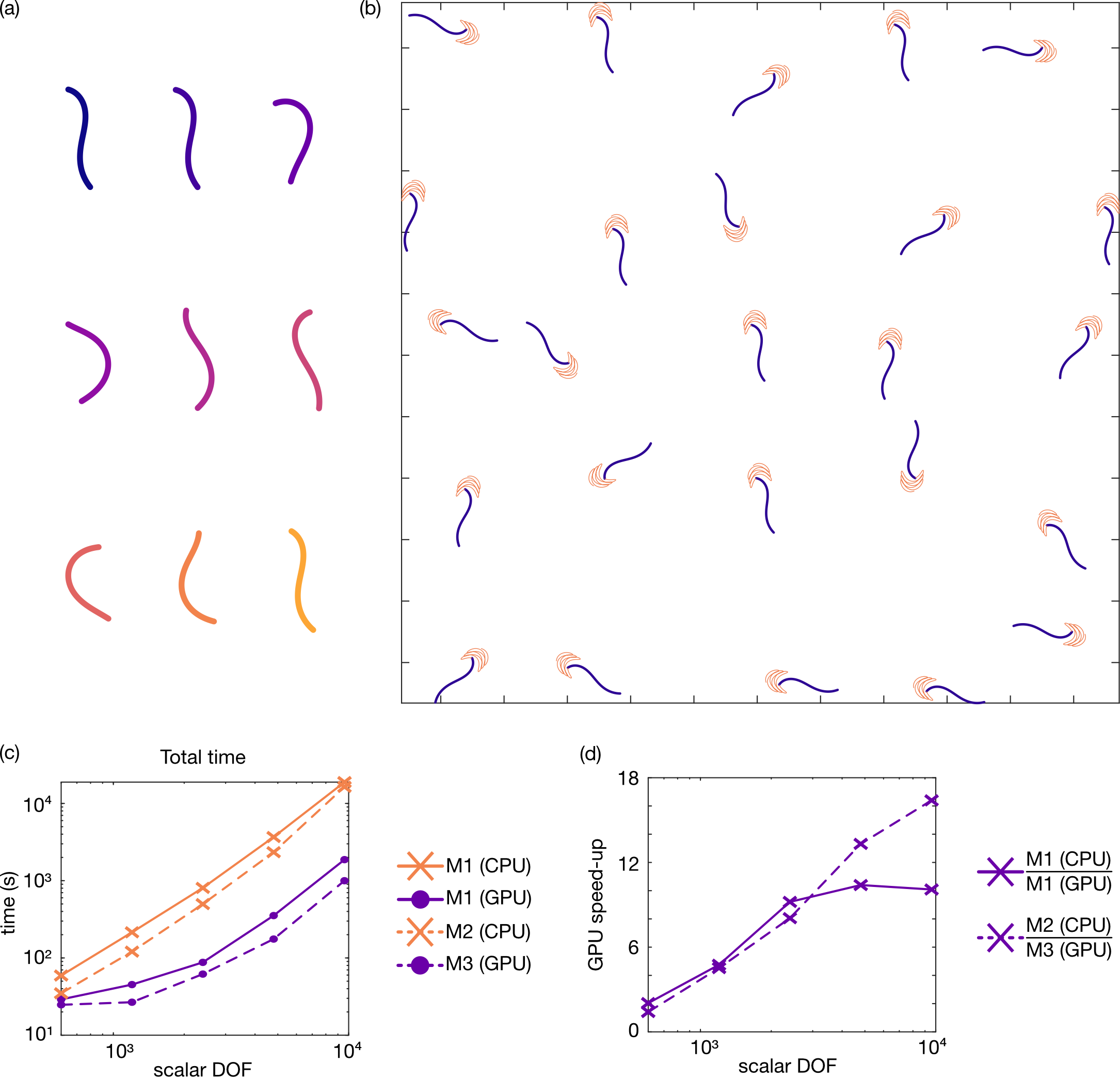}
        \phantomsubcaption{}
        \label{fig:manycelegans_beat}
    \end{subfigure}
    \begin{subfigure}{0\textwidth}
        \phantomsubcaption{}
        \label{fig:manycelegans_tracks}
    \end{subfigure}
    \begin{subfigure}{0\textwidth}
        \phantomsubcaption{}
        \label{fig:manycelegans_time}
    \end{subfigure}
    \begin{subfigure}{0\textwidth}
        \phantomsubcaption{}
        \label{fig:manycelegans_speedup}
    \end{subfigure}
    \caption{Simulation of an array of 25 model \textit{C. elegans} for three beat cycles. (a) The model beat pattern in time. (b) The position of each swimmer at \(t=0\) is shown, with the paths traced out by the leading point as they swim overlain. (c) The total simulation time required on each of M1 (CPU), M2 (GPU), M3 and M4 (lower is better). (d) The relative speed when using a GPU compared to CPU on M1 and M3 compared to M2 (higher is better).}
    \label{fig:manycelegans}
\end{figure}

\section{Parallelisation enables simulation of large problems}

We have shown that the inclusion of GPU parallelisation enables significant speed up for computations using the NEAREST method, both in individual benchmarks (section 4\ref{sec:benchmarks}) and when comparing to `real-world' calculations (sections \ref{sec:multisperm}, \ref{sec:node}). We will now highlight how this method can be used to investigate other problems in computational biology, such how fluid mixing and transport can occur due to the presence of suspensions of microswimmers \cite{leptos2009dynamics,burkholder2017tracer,aragones2018diffusion}. Specifically, we will focus on the simulation of a large array of \textit{C. elegans} swimmers, and the question of whether the small, rapidly decaying, velocity disturbance caused by swimming sperm is sufficient for particle transport.

\subsection{Multiple swimming \textit{C. elegans}}

We simulate an array of \(25\) model \celegans\ over three beat cycles (see Figure~\ref{fig:manycelegans_tracks}). The swimmers are discretised as a line of stokeslets with \(75N\) sDOF, and corresponding \(Q = 25\times 4N\) vector quadrature points. The \celegans\ beat pattern follows that of Thomases and Guy~\cite{thomases2014mechanisms}, illustrated for a single swimmer in Figure~\ref{fig:manycelegans_beat}. The time taken to solve this swimming problem when \(N = 2^n\) \((n=3,4,\hdots,7)\) is shown in Figure~\ref{fig:manycelegans_time}, with the speed up when using the GPU on M1, or using M3 instead of M2, shown in Figure~\ref{fig:manycelegans_speedup}. As with the single iteration calculations in \S4\ref{sec:benchmarks}, when using a reasonable number of points (\(> 96\)  sDOF per swimmer) we see an order of magnitude speed up in calculations. When using \(384\) sDOF to discretize each swimmer, the real-world time required is just over \(4.5\) hours when using the CPU of M2, reducing down to around \(15\) minutes on the GPU of M3.

\begin{figure}[t]
    \centering 
    \begin{subfigure}{\textwidth}
        \includegraphics[width=\textwidth]{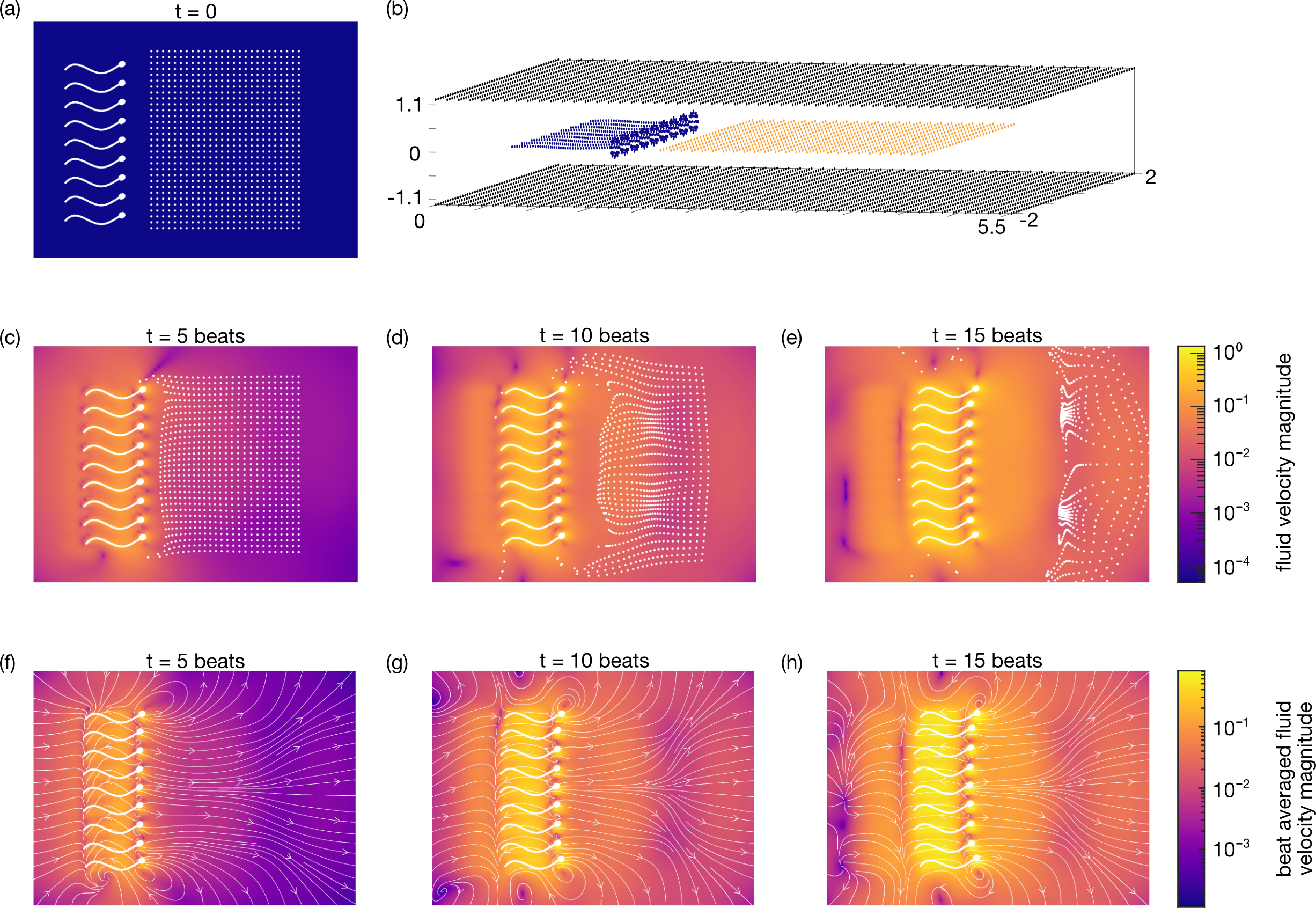}
        \phantomsubcaption{}
        \label{fig:spermentrainment0}
    \end{subfigure}
    \begin{subfigure}{0\textwidth}
        \phantomsubcaption{}
        \label{fig:spermentrainmentsketch}
    \end{subfigure}
    \begin{subfigure}{0\textwidth}
        \phantomsubcaption{}
        \label{fig:spermentrainment5}
    \end{subfigure}
    \begin{subfigure}{0\textwidth}
        \phantomsubcaption{}
        \label{fig:spermentrainment10}
    \end{subfigure}
    \begin{subfigure}{0\textwidth}
        \phantomsubcaption{}
        \label{fig:spermentrainment15}
    \end{subfigure}
    \begin{subfigure}{0\textwidth}
        \phantomsubcaption{}
        \label{fig:spermentrainment5avg}
    \end{subfigure}
    \begin{subfigure}{0\textwidth}
        \phantomsubcaption{}
        \label{fig:spermentrainment10avg}
    \end{subfigure}
    \begin{subfigure}{0\textwidth}
        \phantomsubcaption{}
        \label{fig:spermentrainment15avg}
    \end{subfigure}
    \caption{Simulation of particle transport by a line of 9 model sperm between two flat plates over \(15\) beat cycles. (a) The position of the cells and particles at time \(t=0\). (b) Sketch of the set up showing the cells between two flat plates (note the axes are presented with non-equal scaling, causing the cell heads to appear more circular in this plot than they are in simulations). (c) - (e) The position of the cells and particles at intervals of \(5\) beat cycles, plotted over the velocity magnitude of the fluid disturbance. (f) - (h) The position of the cells at intervals of \(5\) beat cycles, plotted over the beat-averaged velocity magnitude of the fluid disturbance with accompanying `instantaneous' streamlines. In panels (c) - (h) Velocities are scaled with respect to the flagellar length multiplied by the beat frequency, with the colour shown on a logarithmic scale.
    }
    \label{fig:spermentrainment}
\end{figure}

\subsection{Particle transport by sperm swimming between boundaries}

The GPU implementation enables several aspects of interest to be combined to obtain results that may not have previously been possible. In this, final, example we investigate the particle transport by sperm swimming between two flat parallel plates. In this simulation we initialise \(9\) individually modelled sperm, placed in a line aligned half way between two parallel plates, with the centroid of each cell separated by a distance of \(9\ r_2\) (with \(r_2\) being the half-width of the cell head). The rigid boundaries are rectangular with dimensions \(5.5\times 4\) flagellar lengths, and are placed a dimensional distance of \(10~\mu\)m apart to mirror those the common clinical research set-up of a shallow glass imaging chamber. An array of \(26\times 31\) particles, separated in each direction by \(0.1\) flagellar lengths, is placed \(0.5\) flagellar lengths ahead of the sperm. In these simulations the boundaries comprise \(15606\) sDOF, with \(2538\) sDOF for the cells, and \(2418\) sDOF for the particles. A sketch of the cells and particles can be seen as a top-down view in Figure~\ref{fig:spermentrainment0}, and of the cells, particles, and boundaries in Figure~\ref{fig:spermentrainmentsketch}.

The results of these simulations after \(15\) beats are shown in Figure~\ref{fig:spermentrainment}. In panels \ref{fig:spermentrainment5}-\ref{fig:spermentrainment15}, the position of the cells and particles is shown, together with the magnitude of the fluid velocity disturbance (shown on a logarithmic scale). Additionally in panels \ref{fig:spermentrainment5avg}-\ref{fig:spermentrainment15avg}, we show the beat-averaged fluid velocity magnitude, together with `instantaneous' streamlines. A counter-intuitive finding is that these small instantaneous velocities resulting from the rapid decay of the flow field around the swimmers (and even smaller beat-averaged velocities) are nevertheless sufficient to result in transport that, initially at least, has a larger average velocity than the progressive speed of the individual sperm cells driving the flow. The ability to transport particles ahead of the sperm may serve as part of an important biological process; could this provide evidence for a potential mechanism by which a population of sperm can introduce chemical messengers from semen into the cervix and uterus? Further experimental work would be required to confirm this, however the results in Figure~\ref{fig:spermentrainment} suggest it is at least hydrodynamically feasible.

The fluid dynamics of transport in narrow films or between plates is of longstanding interest to the biofluids community. Liron and Mochon \cite{liron1976} showed that the flow generated by a stokeslet between two parallel flat plates (later extended to thin films by Mathijssen \textit{et. al} \cite{mathijssen2016hydrodynamics}) exhibits source-dipole-like far-field behaviour, decaying as \(1 / r^3\). We would hence expect the flow generated by a swimmer (or set of swimmers) to decay like \(1/r^4\); an observation which is consistent with the results in Figure~\ref{fig:spermentrainment}, in which the transported particles are most densely located approximately \(1\) flagellar length away in front of the swimming cells, with the rapid decay causing the particles to bunch together. The problem of particle mixing in such confined flows is also important for understanding many biological processes. Pushkin and Yeomans \cite{pushkin2014stirring} demonstrated how domain confinement significantly changes the transport of particles by swimming cells and its dependence on swimmer kinematics. To complement analysis based on statistical description and idealised far fields, parallel simulation approaches such as those described here should enable more specific details of cell morphology, beat kinematics and geometry/complexity of the confining environment to be taken into account.

The results presented here suggest that GPU parallelisation can enable flow fields and particle disturbance in confined geometries to be precisely resolved; moreover the computational framework should also enable swimming and transport in complex and dynamic environments such as the female reproductive tract to be computed.

\section{Discussion}
The regularized stokeslet method provides a relatively elementary access point to numerical simulation of the geometrically complex Stokes flows characterising many microscale biological systems. In this paper we reviewed the nearest-neighbour discretisation approach, which decouples the number of degrees of freedom from the quadrature process, hence controlling the size of the linear system that needs to be solved. We then described the implementation of this method on GPU-enabled hardware; the fact that the method is already based on linear algebra operations means that only two lines of MATLAB code needed to be added in order to exploit GPU acceleration. Assessing the method on existing problems of calculating swimming motion due to multiple sperm in a confined channel, and particle transport in a ciliated organ, as well as new problems involving multiple undulatory swimmers in an infinite fluid, we consistently observed \textit{at least} an order of magnitude time reduction when using the GPU over the CPU. We have also demonstrated the versatility of this method by assessing the problem of predicting particle transport by multiple sperm swimming in a directed fashion between two parallel plates. While we carried out some of the computations on an HPC cluster, we limited our use to a single computational node per simulation. Further advances may be possible by re-analysing the code and/or utilising a compiler, although our primary focus is simplicity of implementation.

Parallel computing methods for microscale flow have been explored since the development of the completed double layer boundary integral equation method of Phan-Thien and Tullock in 1994 \cite{phan1994} (see also the earlier work of Power and Miranda \cite{power1987} and later by Keaveny \& Shelley \cite{keaveny2011}), enabling a multiparticle system with \(81,000\) degrees of freedom to be solved on a Cray-YMP class supercomputer. The most powerful approaches to large scale microscale flow problems are based on approximating the multipole expansion of the stokeslet, including the kernel-independent fast multipole method \cite{ying2004,rostami2016,nazockdast2017,rostami2019}, treecode \cite{wang2018} and the force-coupling method \cite{delmotte2015,schoeller2018}, which enable \(O(N)\) or \(O(N\log N)\) computational complexity, by contrast with the \(O(N^3)\) complexity of the method described here. These ideas have been used to simulate large suspensions of swimmers and elastic fibres, which are beyond the scope of the `original' formulations of the boundary integral method or regularized stokeslet method, arguably at the cost of significantly greater implementational complexity. More `mainstream' computational fluid dynamics approaches based on volumetric meshes and the finite volume and finite element methods have also been deployed very successfully for multicilia simulations for example \cite{mitran2007}. It would certainly be interesting to compare the present code with these methods, however the spirit of the nearest-neighbour approach is, rather than aiming to reduce asymptotic complexity, to minimise unnecessary degrees of freedom (i.e.\ to reduce \(N\)), via algorithms that can be implemented in a few linear algebra operations, and without intricate mesh construction. Future efforts along these lines may consider how rapidly-advancing algorithmic and hardware developments can continue to be made available to the non-specialist community.

\subsection*{Data access}
MATLAB code to calculate the results and create the figures in this paper can be found at \verb@https://gitlab.com/meuriggallagher/passively-parallel-regularized-stokeslets@, and the GPU accelerated NEAREST package can be found at \verb@https://gitlab.com/meuriggallagher@ \\\verb@/nearest@

\subsection*{Author contributions}
MTG developed the computational implementation and performed the experiments. Both authors devised the research, analysed results and drafted the manuscript.

\subsection*{Funding}
The authors gratefully acknowledge funding from the Engineering and Physical Sciences Research Council (EPSRC) Healthcare Technologies Award (EP/N021096/1). M.T.G. gratefully acknowledges support of the University of Birmingham through its Dynamic Investment Fund.

\subsection*{Acknowledgements}
The authors thank the organisers of the \emph{Stokes at 200} meeting at Pembroke College, Cambridge in September 2019 and acknowledge interesting discussions with other participants that contributed to the exposition. The authors also acknowledge the ongoing contributions of their collaborators, in particular Tom Montenegro-Johnson and Jackson Kirkman-Brown (Birmingham), and thank David Gagnon (Georgetown) for discussions about modelling \textit{C. Elegans}. A significant portion of the computational work described in this paper was performed using the University of Birmingham's BlueBEAR HPC service, which provides a High Performance Computing service to the University's research community. See \verb@http://www.birmingham.ac.uk/bear@ for more details.


\begin{thebibliography}{10}
\expandafter\ifx\csname urlstyle\endcsname\relax
  \providecommand{\doi}[1]{(doi:\discretionary{}{}{}#1)}\else
  \providecommand{\doi}{(doi:\discretionary{}{}{}\begingroup
  \urlstyle{rm}\Url)}\fi

\bibitem{stokes1851}
Stokes G. 1851 On the effect of the internal friction of fluids on the motion
  of pendulums.
\newblock \emph{Trans. Camb. Phil. Soc.} \textbf{9}, 8--106.

\bibitem{gray1955}
Gray J, Hancock G. 1955 The propulsion of sea-urchin spermatozoa.
\newblock \emph{J. Exp. Biol.} \textbf{32}, 802--814.

\bibitem{hancock1953}
Hancock G. 1953 The self-propulsion of microscopic organisms through liquids.
\newblock \emph{Proc. R. Soc. Lond. A} \textbf{217}, 96--121.

\bibitem{burgers1938}
Burgers J. 1938 On the motion of small particles of elongated form suspended in
  a viscous liquid.
\newblock \emph{Kon. Ned. Akad. Wet. Verhand. (Eerste Sectie)} \textbf{16},
  113--184.

\bibitem{lighthill1976}
Lighthill J. 1976 Flagellar hydrodynamics.
\newblock \emph{SIAM Rev.} \textbf{18}, 161--230.

\bibitem{johnson1980}
Johnson R. 1980 An improved slender-body theory for {S}tokes flow.
\newblock \emph{J. Fluid Mech.} \textbf{99}, 411--431.

\bibitem{chwang1971}
Chwang A, Wu T. 1971 A note on the helical movement of micro-organisms.
\newblock \emph{Proc. R. Soc. Lond. Ser. B.} \textbf{178}, 327--346.

\bibitem{ramia1993}
Ramia M, Tullock D, Phan-Thien N. 1993 The role of hydrodynamic interaction in
  the locomotion of microorganisms.
\newblock \emph{Biophys. J.} \textbf{65}, 755--778.

\bibitem{lauga2006}
Lauga E, DiLuzio W, Whitesides G, Stone H. 2006 Swimming in circles: motion of
  bacteria near solid boundaries.
\newblock \emph{Biophys. J.} \textbf{90}, 400--412.

\bibitem{blake1971squirmer}
Blake J. 1971 {Self propulsion due to oscillations on the surface of a cylinder
  at low Reynolds number}.
\newblock \emph{Bull. Austr. Math. Soc.} \textbf{5}, 255--264.

\bibitem{blake1971images}
Blake J. 1971 A note on the image system for a stokeslet in a no-slip boundary.
\newblock \emph{Proc. Camb. Phil. Soc.} \textbf{70}, 303--310.

\bibitem{ishimoto2013}
Ishimoto K. 2013 {A spherical squirming swimmer in unsteady Stokes flow}.
\newblock \emph{J. Fluid Mech.} \textbf{723}, 163--189.

\bibitem{pedley2016}
Pedley T, Brumley D, Goldstein R. 2016 {Squirmers with swirl: a model for
  Volvox swimming}.
\newblock \emph{J. Fluid Mech.} \textbf{798}, 165--186.

\bibitem{liron1976}
Liron N, Mochon S. 1976 Stokes flow for a stokeslet between two parallel flat
  plates.
\newblock \emph{J. Eng. Math.} \textbf{10}, 287--303.

\bibitem{blake1996}
Blake J, Otto S. 1996 Ciliary propulsion, chaotic filtration and a `blinking'
  stokeslet.
\newblock \emph{J. Eng. Math.} \textbf{30}, 151--168.

\bibitem{ainley2008}
Ainley J, Durkin S, Embid R, Boindala P, Cortez R. 2008 {The method of images
  for regularized Stokeslets}.
\newblock \emph{J. Comp. Phys.} \textbf{227}, 4600--4616.

\bibitem{cortez2015}
Cortez R, Varela D. 2015 A general system of images for regularized
  {S}tokeslets and other elements near a plane wall.
\newblock \emph{J. Comp. Phys.} \textbf{285}, 41--54.

\bibitem{pedley1987}
Pedley T, Kessler J. 1987 The orientation of spheroidal microorganisms swimming
  in a flow field.
\newblock \emph{Proc. R. Soc. Lond. Ser. B.} \textbf{231}, 47--70.

\bibitem{hill2005}
Hill N, Pedley T. 2005 Bioconvection.
\newblock \emph{Fluid Dyn. Res.} \textbf{37}, 1.

\bibitem{cartwright2004}
Cartwright J, Piro O, Tuval I. 2004 Fluid-dynamical basis of the embryonic
  development of left-right asymmetry in vertebrates.
\newblock \emph{Proc. Natl. Acad. Sci.} \textbf{101}, 7234--7239.

\bibitem{brokaw2005}
Brokaw C. 2005 {Computer simulation of flagellar movement IX. Oscillation and
  symmetry breaking in a model for short flagella and nodal cilia}.
\newblock \emph{Cell Motil. Cytoskel.} \textbf{60}, 35--47.

\bibitem{cartwright2020}
Cartwright J, Piro O, Tuval I. 2020 Chemosensing versus mechanosensing in nodal
  and {K}upffer's vesicle cilia and in other left-right organizer organs.
\newblock \emph{Phil. Trans. R. Soc. Lond. Ser. B} \textbf{375}, 20190566.

\bibitem{cortez2001}
Cortez R. 2001 The method of regularized {S}tokeslets.
\newblock \emph{SIAM J. Sci. Comput.} \textbf{23}, 1204--1225.

\bibitem{cortez2005}
Cortez R, Fauci L, Medovikov A. 2005 The method of regularized {S}tokeslets in
  three dimensions: analysis, validation, and application to helical swimming.
\newblock \emph{Phys. Fluids} \textbf{17}, 031504.

\bibitem{goldstein2015}
Goldstein R. 2015 Green algae as model organisms for biological fluid dynamics.
\newblock \emph{Annu. Rev. Fluid Mech.} \textbf{47}, 343--375.

\bibitem{lauga2009}
Lauga E, Powers T. 2009 The hydrodynamics of swimming microorganisms.
\newblock \emph{Rep. Progr. Phys.} \textbf{72}, 096601.

\bibitem{montenegro2012}
Montenegro-Johnson T, Smith A, Smith D, Loghin D, Blake J. 2012 Modelling the
  fluid mechanics of cilia and flagella in reproduction and development.
\newblock \emph{Eur. Phys. J. E} \textbf{35}, 111.

\bibitem{elgeti2015}
Elgeti J, Winkler R, Gompper G. 2015 Physics of microswimmers--single particle
  motion and collective behavior: a review.
\newblock \emph{Rep. Progr. Phys.} \textbf{78}, 056601.

\bibitem{goldstein2016}
Goldstein R. 2016 Batchelor prize lecture fluid dynamics at the scale of the
  cell.
\newblock \emph{J. Fluid Mech.} \textbf{807}, 1--39.

\bibitem{smith2019}
Smith D, Montenegro-Johnson T, Lopes S. 2019 Symmetry-breaking cilia-driven
  flow in embryogenesis.
\newblock \emph{Annu. Rev. Fluid Mech.} \textbf{51}, 105--128.

\bibitem{gallagher2018}
Gallagher M, Smith D. 2018 Meshfree and efficient modeling of swimming cells.
\newblock \emph{Phys. Rev. Fluids} \textbf{3}, 053101.

\bibitem{gallagher2020}
Gallagher M, Montenegro-Johnson T, Smith D. 2020 Simulations of particle
  tracking in the oligociliated mouse node and implications for left--right
  symmetry-breaking mechanics.
\newblock \emph{Phil. Trans. R. Soc. Ser. B.} \textbf{375}, 20190161.

\bibitem{purcell1977}
Purcell E. 1977 Life at low {R}eynolds number.
\newblock \emph{Amer. J. Phys.} \textbf{45}, 3--11.

\bibitem{taylor1985}
Taylor G. 1985 \emph{Low Reynolds number flows}.
\newblock Released by Encyclopaedia Britannica Educational Corporation.

\bibitem{oseen1927}
Oseen CW. 1927 Neuere methoden und ergebnisse in der hydrodynamik.
\newblock \emph{Leipzig: Akademische Verlagsgesellschaft mb H.} .

\bibitem{youngren1975}
Youngren G, Acrivos A. 1975 Stokes flow past a particle of arbitrary shape: a
  numerical method of solution.
\newblock \emph{J. Fluid Mech.} \textbf{69}, 377--403.

\bibitem{phan1987}
Phan-Thien N, Tran-Cong T, Ramia M. 1987 A boundary-element analysis of
  flagellar propulsion.
\newblock \emph{J. Fluid Mech.} \textbf{184}, 533--549.

\bibitem{pozrikidis1992}
Pozrikidis C. 1992 \emph{Boundary integral and singularity methods for
  linearized viscous flow}.
\newblock Cambridge University Press.

\bibitem{ishimoto2017}
Ishimoto K, Gaffney E. 2017 Boundary element methods for particles and
  microswimmers in a linear viscoelastic fluid.
\newblock \emph{J. Fluid Mech.} \textbf{831}, 228--251.

\bibitem{pozrikidis2002}
Pozrikidis C. 2002 \emph{A practical guide to boundary element methods with the
  software library {BEMLIB}}.
\newblock CRC Press.

\bibitem{nystrom1930}
Nystr{\"o}m E. 1930 {\"U}ber die praktische aufl{\"o}sung von
  integralgleichungen mit anwendungen auf randwertaufgaben.
\newblock \emph{Acta Math.} \textbf{54}, 185--204.

\bibitem{gallagher2019}
Gallagher M, Choudhuri D, Smith D. 2019 Sharp quadrature error bounds for the
  nearest-neighbor discretization of the regularized stokeslet boundary
  integral equation.
\newblock \emph{SIAM J. Sci. Comput.} \textbf{41}, B139--B152.

\bibitem{smith2012}
Smith A, Johnson T, Smith D, Blake J. 2012 Symmetry breaking cilia-driven flow
  in the zebrafish embryo.
\newblock \emph{J. Fluid Mech.} \textbf{705}, 26--45.

\bibitem{sampaio2014}
Sampaio P, Ferreira R, Guerrero A, Pintado P, Tavares B, Amaro J, Smith A,
  Montenegro-Johnson T, Smith D, Lopes S. 2014 Left-right organizer flow
  dynamics: how much cilia activity reliably yields laterality?
\newblock \emph{Dev. Cell} \textbf{29}, 716--728.

\bibitem{montenegro2015}
Montenegro-Johnson T, Michelin S, Lauga E. 2015 A regularised singularity
  approach to phoretic problems.
\newblock \emph{Eur. Phys. J. E} \textbf{38}, 139.

\bibitem{schuech2019}
Schuech R, Hoehfurtner T, Smith D, Humphries S. 2019 Motile curved bacteria are
  {P}areto-optimal.
\newblock \emph{Proc. Natl. Acad. Sci.} \textbf{116}, 14440--14447.

\bibitem{smith2009}
Smith D. 2009 A boundary element regularized {S}tokeslet method applied to
  cilia-and flagella-driven flow.
\newblock \emph{Proc. R. Soc. Lond. Ser. A} \textbf{465}, 3605--3626.

\bibitem{cortez2018}
Cortez R. 2018 Regularized stokeslet segments.
\newblock \emph{J. Comput. Phys.} \textbf{375}, 783--796.

\bibitem{walker2019}
Walker B, Ishimoto K, Gadêlha H, Gaffney E. 2019 Filament mechanics in a
  half-space via regularised {S}tokeslet segments.
\newblock \emph{J. Fluid Mech.} \textbf{879}, 808–833.

\bibitem{hall2019}
Hall-McNair A, Montenegro-Johnson T, Gad{\^e}lha H, Smith D, Gallagher M. 2019
  Efficient implementation of elastohydrodynamics via integral operators.
\newblock \emph{Phys. Rev. Fluids} \textbf{4}, 113101.

\bibitem{smith2018}
Smith DJ. 2018 A nearest-neighbour discretisation of the regularized stokeslet
  boundary integral equation.
\newblock \emph{J. Comput. Phys.} \textbf{358}, 88--102.

\bibitem{rostami2016}
Rostami M, Olson S. 2016 Kernel-independent fast multipole method within the
  framework of regularized {S}tokeslets.
\newblock \emph{J. Fluid. Struct.} \textbf{67}, 60--84.

\bibitem{rostami2019}
Rostami M, Olson S. 2019 Fast algorithms for large dense matrices with
  applications to biofluids.
\newblock \emph{J. Comput. Phys.} .

\bibitem{wang2018}
Wang L, Tlupova S, Krasny R. 2018 A treecode algorithm for 3{D} {S}tokeslets
  and stresslets.
\newblock \emph{arXiv preprint arXiv:1811.12498} .

\bibitem{dziekonski2012finite}
Dziekonski A, Sypek P, Lamecki A, Mrozowski M. 2012 Finite element matrix
  generation on a gpu.
\newblock \emph{Pr. Electromagn. Res.} \textbf{128}, 249--265.

\bibitem{dresdner1981relationships}
Dresdner R, Katz D. 1981 Relationships of mammalian sperm motility and
  morphology to hydrodynamic aspects of cell function.
\newblock \emph{Biol. Reprod.} \textbf{25}, 920--930.

\bibitem{leptos2009dynamics}
Leptos K, Guasto J, Gollub J, Pesci A, Goldstein R. 2009 Dynamics of enhanced
  tracer diffusion in suspensions of swimming eukaryotic microorganisms.
\newblock \emph{Phys. Rev. Lett.} \textbf{103}, 198103.

\bibitem{burkholder2017tracer}
Burkholder E, Brady J. 2017 Tracer diffusion in active suspensions.
\newblock \emph{Phys. Rev. E} \textbf{95}, 052605.

\bibitem{aragones2018diffusion}
Aragones J, Yazdi S, Alexander-Katz A. 2018 Diffusion of self-propelled
  particles in complex media.
\newblock \emph{Phys. Rev. Fluids} \textbf{3}, 083301.

\bibitem{thomases2014mechanisms}
Thomases B, Guy R. 2014 Mechanisms of elastic enhancement and hindrance for
  finite-length undulatory swimmers in viscoelastic fluids.
\newblock \emph{Phys. Rev. Lett.} \textbf{113}, 098102.

\bibitem{mathijssen2016hydrodynamics}
Mathijssen A, Doostmohammadi A, Yeomans J, Shendruk T. 2016 Hydrodynamics of
  micro-swimmers in films.
\newblock \emph{J. Fluid Mech.} \textbf{806}, 35--70.

\bibitem{pushkin2014stirring}
Pushkin D, Yeomans J. 2014 Stirring by swimmers in confined microenvironments.
\newblock \emph{J. Stat. Mech.-Theory E.} \textbf{2014}, P04030.

\bibitem{phan1994}
Phan-Thien N, Tullock D. 1994 {Completed double layer boundary element method
  in elasticity and stokes flow: Distributed computing through PVM}.
\newblock \emph{Comput. Mech.} \textbf{14}, 370--383.

\bibitem{power1987}
Power H, Miranda G. 1987 {Second kind integral equation formulation of
  Stokes’ flows past a particle of arbitrary shape}.
\newblock \emph{SIAM J. Appl. Math.} \textbf{47}, 689--698.

\bibitem{keaveny2011}
Keaveny E, Shelley M. 2011 {Applying a second-kind boundary integral equation
  for surface tractions in Stokes flow}.
\newblock \emph{J. Comput. Phys.} \textbf{230}, 2141--2159.

\bibitem{ying2004}
Ying L, Biros G, Zorin D. 2004 A kernel-independent adaptive fast multipole
  algorithm in two and three dimensions.
\newblock \emph{J. Comput. Phys.} \textbf{196}, 591--626.

\bibitem{nazockdast2017}
Nazockdast E, Rahimian A, Zorin D, Shelley M. 2017 A fast platform for
  simulating semi-flexible fiber suspensions applied to cell mechanics.
\newblock \emph{J. Comput. Phys.} \textbf{329}, 173--209.

\bibitem{delmotte2015}
Delmotte B, Keaveny E, Plourabou{\'e} F, Climent E. 2015 Large-scale simulation
  of steady and time-dependent active suspensions with the force-coupling
  method.
\newblock \emph{J. Comput. Phys.} \textbf{302}, 524--547.

\bibitem{schoeller2018}
Schoeller S, Keaveny E. 2018 From flagellar undulations to collective motion:
  predicting the dynamics of sperm suspensions.
\newblock \emph{J. R. Soc. Interface} \textbf{15}, 20170834.

\bibitem{mitran2007}
Mitran S. 2007 Metachronal wave formation in a model of pulmonary cilia.
\newblock \emph{Comput. Struct.} \textbf{85}, 763--774.

\end{thebibliography}
\end{document}